\documentclass[conference]{IEEEtran}

\usepackage[T1]{fontenc}
\usepackage{amsmath}
\usepackage{amssymb}
\usepackage{stmaryrd}
\usepackage{epic,eepic}
\usepackage{theorem}
\usepackage{pifont}  
\usepackage{euscript}
\usepackage{calc}
\usepackage[titles]{tocloft}

\usepackage[usenames]{color}          
\usepackage{xcolor}
\definecolor{Brown}{rgb}{0.55,0.0,0.10}
\definecolor{dgreen}{rgb}{0.00,0.56,0.00}
\definecolor{vertmoinsfonce}{rgb}{0.00,0.50,0.00}
\definecolor{vert}{rgb}{0.00,0.60,0.00}
\definecolor{llightggray}{rgb}{0.97,0.97,0.97}
\definecolor{lightggray}{rgb}{0.9,0.9,0.9}
\definecolor{ggray}{rgb}{0.5,0.5,0.5}
\definecolor{darkggray}{rgb}{0.25,0.25,0.25}
\definecolor{ddarkggray}{rgb}{0.1,0.1,0.1}
\definecolor{bleu}{rgb}{0.00,0.00,1.00}

\usepackage{epsf}           
\usepackage{graphicx}       
\usepackage{epsfig}         
\usepackage{pstricks}
\usepackage{pstricks-add}
\usepackage{pst-plot}
\usepackage{dsfont}
\theoremheaderfont{\color{black}\normalfont\bfseries}

\newtheorem{theorem}{Theorem}[section]
\newtheorem{definition}[theorem]{Definition}
\newtheorem{corollary}[theorem]{Corollary}

\theoremstyle{plain}{\theorembodyfont{\rmfamily}%
}
\theoremstyle{plain}{\theorembodyfont{\rmfamily}%
}
\theoremstyle{plain}{
\theorembodyfont{\rmfamily}

	\newtheorem{remark}[theorem]{Remark}
	
	}

\newcommand{\R}{\mathbb{R}}

\newcommand{\N}{\mathbb{N}}

\newcommand{\E}{\mathbb{E}}
\newcommand{\Q}{\mathbb{Q}}

\font\dsrom=dsrom10 scaled 1200 \def \indic{\textrm{\dsrom{1}}}
\newcommand{\UN}{\indic}

\newcommand{\C}{\mathcal{C}}

\newcommand{\QQ}{\mathcal{Q}}

\newcommand{\D}{\mathcal{D}}

\newcommand{\PP}{\mathcal{P}}

\newcommand{\mc}{\mathcal}

{\Large\normalsize}
{\Large\normalsize}

\ifCLASSINFOpdf
\else
\fi
\begin{document}

\title{Empirical Coordination with Two-Sided State Information and Correlated Source and State}

%
%
%

\author{\IEEEauthorblockN{Ma\"{e}l Le Treust}\\
\IEEEauthorblockA{
ETIS, UMR 8051 / ENSEA, Université Cergy-Pontoise, CNRS,\\
6, avenue du Ponceau,\\
95014 CERGY-PONTOISE CEDEX,\\
FRANCE\\
Email: mael.le-treust@ensea.fr}
}



\maketitle

\begin{abstract}

The coordination of autonomous agents is a critical issue for decentralized communication networks. Instead of transmitting information, the agents interact in a coordinated manner in order to optimize a general objective function. A target joint probability distribution is achievable if there exists a code such that the sequences of symbols are jointly typical. The empirical coordination is strongly related to the joint source-channel coding with two-sided state information and correlated source and state. This problem is also connected to state communication and is open for non-causal encoder and decoder. We characterize the optimal solutions for perfect channel, for lossless decoding, for independent source and channel, for causal encoding and for causal decoding.


\end{abstract}


%
\IEEEpeerreviewmaketitle

\section{Introduction}\label{sec:Introduction}


The problem of the coordination of autonomous agents is the cornerstone of decentralized communication networks. Communication devices are considered as agents that interact in a coordinated manner in order to achieve a common objective, for example, the transmission of information. This analysis is based on a two step approach \cite{AnantharamBorkar07}, \cite{LeTreust(EmpiricalCoordination)14}. The first step is the characterization of the set of achievable joint probability distributions over the symbols of source and channel. The second step is the maximization or the minimization of an objective function (source distortion or channel cost) by considering the set of achievable joint probability distributions.

Empirical coordination has been investigated in \cite{KramerSavari07}, \cite{CuffPermuterCover10} with a rate-distortion perspective. In the literature of game theory, the agents coordinate their actions by implementing a coding scheme that satisfies an information constraint \cite{GoHerNey06}. Empirical coordination was under investigation in \cite{Cuff(ImplicitCoordination)10} for perfect channel, in \cite{CuffSchieler11} for causal encoding, in \cite{LeTreust(ISITfeedbacks)15} with channel feedback, in \cite{LetreustZaidiLasaulce(Allerton)11} for a multi-user network with an eavesdropper, in \cite{BlochLuzziKliewer12} for polar codes, 
in \cite{LeTreust(EmpiricalCoordination)14}  for causal decoding, in \cite{SamsonBenjaminISIT13}, \cite{LarrousseLasaulceBloch(IT)14} with feedback from the source, and in \cite{LeTreust(CorrelationITW)14} for lossless decoding and correlated source and state.

\begin{figure}[!ht]
\begin{center}
\psset{xunit=0.9cm,yunit=0.9cm}
\begin{pspicture}(0,-0.35)(8.5,1.7)
\pscircle(0,0.5){0.45}
\psframe(2,0)(3,1)
\pscircle(5,0.5){0.45}
\psframe(7,0)(8,1)
\psline[linewidth=1pt]{->}(0.5,0.5)(2,0.5)
\psline[linewidth=1pt]{->}(3,0.5)(4.5,0.5)
\psline[linewidth=1pt]{->}(5.5,0.5)(7,0.5)
\psline[linewidth=1pt]{->}(8,0.5)(9,0.5)
\psline[linewidth=1pt]{->}(0,0)(0,-0.5)(5,-0.5)(5,0)
\psline[linewidth=1pt]{->}(2.5,-0.5)(2.5,0)
\psline[linewidth=1pt]{->}(0,1)(0,1.5)(7.5,1.5)(7.5,1)
\rput[u](1,0.8){$U^n$}
\rput[u](1,-0.2){$S^n$}
\rput[u](1,1.8){$Z^{n}$}
\rput[u](3.75,0.8){$X^n$}
\rput[u](6.25,0.8){$Y^{n}$}
\rput[u](8.5,0.8){$V^n$}
\rput(0,0.5){$\PP_{\sf{usz}}$}
\rput(2.5,0.5){$\C$}
\rput(5,0.5){$\mc{T}$}
\rput(7.5,0.5){$\D$}
\end{pspicture}
\caption{Non-causal encoding $f: \mc{U}^n \times \mc{S}^n \rightarrow \mc{X}^n$ and decoding $g: \mc{Y}^n \times \mc{Z}^n \rightarrow \mc{V}^n$  functions for i.i.d. source $\PP_{\sf{usz}}$ and  channel $\mc{T}_{\sf{y|xs}}$. }\label{fig:TwoSided}
\end{center}
\end{figure}
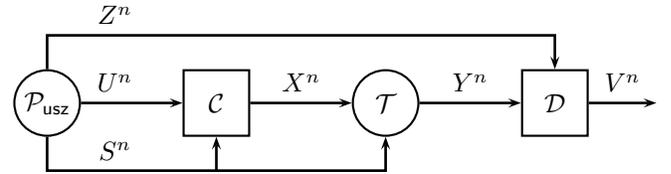

The characterization of the set of achievable joint probability distributions is equivalent to the joint source-channel coding with two-sided state information \cite{MerhavShamai03}, \cite{CoverChiang02} and correlated source and state (Fig. \ref{fig:TwoSided}). It is also related to the problem of state communication \cite{ChoudhuriKimMitra13}. In this paper, we investigate the empirical coordination with non-causal encoding and decoding and we characterize the optimal solutions for three particular cases: perfect channel, lossless decoding and independent source and channel. We also characterize the optimal solutions for causal encoding and causal decoding with two-sided state information.

\vspace{-0.5cm}

Channel model and definitions are presented in Sec. \ref{sec:ModelDefinition}. In Sec. \ref{sec:Achievability}, we provide achievability and converse results for non-causal encoding and decoding. In Sec. \ref{sec:OptimalSolutions},  we characterize optimal solutions for perfect channel, for lossless decoding and for independent source and channel. Causal encoding and decoding are presented in Sec. \ref{sec:CausalEncDec}. Conclusion is in Sec. \ref{sec:Conclusion} and sketches of proof are provided in App. \ref{sec:ProofAchievabilityNC}, \ref{sec:ProofConverseNC}, \ref{sec:ConversePC}, \ref{app:theoCE}, \ref{app:theoCEconv}.

\section{System Model}\label{sec:ModelDefinition}

The problem under investigation is depicted in Fig. \ref{fig:TwoSided}. Capital letter $U$ denotes the random variable, lowercase letter $u\in\mc{U}$ denotes the realization and $\mc{U}^n$ denotes the $n$-time cartesian product. $U^n$, $S^n$, $Z^n$, $X^n$, $Y^n$,  $V^n$ denote the sequences of random variables of source symbols $u^n=(u_1,\ldots,u_n)\in\mc{U}^n$,  of channel states $s^n\in\mc{S}^n$, of state informations at the decoder $z^n\in\mc{Z}^n$, of channel inputs $x^n\in\mc{X}^n$, of channel outputs $y^n\in\mc{Y}^n$ and of outputs of the decoder $v^n\in\mc{V}^n$. Sets $\mc{U}$, $\mc{S}$, $\mc{Z}$, $\mc{X}$, $\mc{Y}$, $\mc{V}$ are discrete. $\Delta(\mc{X})$ stands for the set of  probability distributions $\PP(X)$ over $\mc{X}$. The total variation distance between the probability distributions $\QQ$ and $\PP$ is denoted by $||\QQ - \PP||_{\sf{tv}}= 1/2\cdot \sum_{x\in\mc{X}} |\QQ(x) - \PP(x)|$. Notation $\UN(y|x)$ denotes the indicator function, that is equal to 1 if $y = x$ and 0 otherwise. Markov chain property is denoted by $Y  -\!\!\!\!\minuso\!\!\!\!-X    -\!\!\!\!\minuso\!\!\!\!-  U$ and holds if for all $(u,x,y)$ we have $\PP(y|x,u) = \PP(y|x)$. Information source and channel states are correlated and i.i.d. distributed with $\PP_{\sf{usz}}$. Channel is memoryless  with transition probability $\mc{T}_{\sf{y|xs}}$. Statistics of $\PP_{\sf{usz}}$ and $\mc{T}_{\sf{y|xs}}$ are known by both encoder $\C$ and decoder $\D$.



\begin{definition}\label{def:Code}
Non-causal code $c=(f,g)\in\mc{C}(n)$ is defined by:
\vspace{-0.45cm}
\begin{eqnarray}
&f:&  \mc{U}^n \times \mc{S}^n  \longrightarrow \mc{X}^n ,\label{eq:CodeSource1}\\
&g:&  \mc{Y}^n \times \mc{Z}^n \longrightarrow \mc{V}^n.\label{eq:CodeSource2}
\end{eqnarray}
$\textsf{N}(u|u^n)$ denotes the occurrence number of symbol $u \in \mc{U}$ in the sequence $u^n$. The empirical distribution ${Q}^n \in \Delta(\mc{U}\times \mc{S} \times \mc{Z} \times \mc{X}\times \mc{Y} \times\mc{V})$ of  $(u^n,s^n,z^n,x^n,y^n,v^n)$
is defined by:
\begin{small}
\begin{eqnarray}
{Q}^n(u,s,z,x,y,v) &=& \frac{\textsf{N}(u,s,z,x,y,v |u^n,s^n,z^n,x^n,y^n,v^n)}{n},  \nonumber \\
 \forall  (u,s,z,x,y,v)& \in& \mc{U}\times \mc{S}\times \mc{Z}\times \mc{X} \times \mc{Y} \times\mc{V}. \label{eq:EmpiricalDistribution}
\end{eqnarray}
\end{small}
Fix a target proba. distribution $\QQ \in \Delta(\mc{U} \times \mc{S} \times \mc{Z} \times \mc{X} \times \mc{Y} \times \mc{V}  )$, the error probability of the code $c\in\mc{C}(n)$ is defined by:
\begin{eqnarray}
\PP_{\textsf{e}}(c) = \PP_c\bigg(\Big|\Big|Q^n - \QQ \Big|\Big|_{\sf{tv}}\geq \varepsilon\bigg),\label{eq:ErrorProba}
\end{eqnarray}
where $Q^n \in \Delta(\mc{U}\times \mc{S} \times \mc{Z}\times \mc{X}\times \mc{Y} \times\mc{V})$ is the random variable of the empirical distribution 
induced by the code $c \in \mc{C}(n)$ and the probability distributions $\PP_{\sf{usz}}$, $\mc{T}_{\sf{y|xs}}$.
\end{definition}

\begin{definition}
Probability distribution $\QQ \in  \Delta(\mc{U} \times \mc{S} \times \mc{Z}\times \mc{X} \times \mc{Y} \times \mc{V}  )$ is achievable if for all $\varepsilon>0$, there exists a $\bar{n}\in \N$ s.t. for all $n \geq \bar{n}$, there exists a code $c\in\mc{C}(n)$ that satisfies:
\begin{eqnarray}
\PP_{\textsf{e}}(c)  = \PP_c\bigg(\Big|\Big|Q^n - \QQ \Big|\Big|_{\sf{tv}}\geq \varepsilon\bigg) \leq \varepsilon.
\end{eqnarray}
\end{definition}
If the error probability $\PP_{\textsf{e}}(c)$ is small, the empirical frequency of symbols $(u,s,z,x,y,v) \in \mc{U}\times \mc{S}\times \mc{Z}\times\mc{X}\times \mc{Y} \times\mc{V}$ is close to the target probability distribution $\QQ(u,s,z,x,y,v)$, \textit{i.e.} the sequences $(U^n,S^n,Z^n,X^n,Y^n,V^n)\in A_{\varepsilon}^{{\star}{n}}(\QQ)$ are jointly typical, with large probability. In that case, the sequences of symbols are coordinated empirically. 

The performance of the coordination is evaluated by an objective function  $\Phi : \mc{U} \times \mc{S} \times \mc{Z} \times \mc{X}  \times \mc{Y} \times \mc{V} \mapsto \R$, as stated in \cite{LeTreust(EmpiricalCoordination)14} and \cite{LeTreust(CorrelationITW)14}. This approach is powerful and since the expectation $\E[d(u,v)]$ over the set of achievable distributions, provides directly the minimal distortion level $d(u,v)$. This is valid for any function, as the channel cost $c(x)$ or the utility $\Phi(u,s,z,x,y,v)$ of a decentralized network \cite{GoHerNey06}.



\section{Achievability and Converse Results}\label{sec:Achievability}

We provide necessary and sufficient conditions for non-causal encoding and decoding. This problem is open.

\begin{theorem}[Non-Causal Encoding and Decoding]
$\qquad\qquad\qquad\qquad\qquad\qquad\qquad\qquad\qquad\qquad\qquad$\label{theo:NC}
$1)$ If the joint probability distribution $\QQ(u,s,z,x,y,v)$ is achievable, then it satisfies the marginal and Markov chains:
\begin{small}
\begin{eqnarray}
\begin{cases}
&\QQ(u,s,z) = \PP_{\sf{usz}}(u,s,z),\quad \QQ(y|x,s) = \mc{T}(y | x ,s),\\
&Y -\!\!\!\!\minuso\!\!\!\!- ( X,S)    -\!\!\!\!\minuso\!\!\!\!-   (U, Z) , \quad Z -\!\!\!\!\minuso\!\!\!\!- ( U,S)    -\!\!\!\!\minuso\!\!\!\!-   (X, Y) .
\end{cases}
\end{eqnarray}
\end{small}
2) The joint probability distribution $\PP_{\sf{usz}}(u,s,z)   \otimes   \QQ(x|u,s)  \otimes \mc{T}(y | x ,s)  \otimes \QQ(v|u,s,z,x,y) $ is achievable if:
\begin{eqnarray}
\max_{{\QQ}\in \Q} \bigg( I(W_1,W_2 ; Y, Z  )   -  I( W_1,W_2 ; U, S )   \bigg)  > 0 ,  \label{eq:Achievability1}
\end{eqnarray}
3) The joint probability distribution $\PP_{\sf{usz}}(u,s,z)   \otimes   \QQ(x|u,s)  \otimes \mc{T}(y | x ,s)  \otimes \QQ(v|u,s,z,x,y) $ is not achievable if:
\begin{eqnarray}
\max_{{\QQ}\in \Q} \bigg( I(W_1 ; Y, Z  | W_2 )   -  I( W_2 ; U, S | W_1 )   \bigg)  < 0 ,  \label{eq:Converse1}
\end{eqnarray}
$\Q$ is the set of distributions ${\QQ}   \in  \Delta(\mc{U}   \times \mc{S}  \times \mc{Z}  \times \mc{W}_1\times \mc{W}_2 \times \mc{X} \times \mc{Y}\times \mc{V}  )$ with auxiliary random variables $(W_1,W_2)$ s.t.:
\begin{small}
\begin{eqnarray*}
\begin{cases}
\sum_{(w_1,w_2)\in \mc{W}_1\times \mc{W}_2} {\QQ}(u,s,z,w_1,w_2,x,y,v) \\
\quad =  \PP_{\sf{usz}}(u,s,z)   \otimes   \QQ(x|u,s)\otimes \mc{T}(y | x ,s)   \otimes   \QQ(v|u,s,z,x,y) , \\
Y -\!\!\!\!\minuso\!\!\!\!- (X , S ) -\!\!\!\!\minuso\!\!\!\!-  (U,Z, W_1,W_2 ) ,\\
Z -\!\!\!\!\minuso\!\!\!\!- (U , S ) -\!\!\!\!\minuso\!\!\!\!-  (X , Y, W_1, W_2 ) ,\\
V -\!\!\!\!\minuso\!\!\!\!- (Y,Z,W_1, W_2 ) -\!\!\!\!\minuso\!\!\!\!-  (U,S,X ) .
\end{cases}
\end{eqnarray*}
\end{small}
The probability distribution $\QQ \in \Q$ decomposes as follows:
\begin{small}
\begin{eqnarray*}
\PP_{\sf{usz}}(u,s,z)   \otimes   \QQ(x,w_1,w_2|u,s)  \otimes \mc{T}(y | x ,s)  \otimes \QQ(v|y,z,w_1, w_2).
 \end{eqnarray*}
 \end{small}
The supports of the auxiliary random variables $(W_1,W_2)$ are bounded by $\max(|\mc{W}_1|,|\mc{W}_2|) \leq   (|\mc{B} |  +1) \cdot(|\mc{B} |  +2)$ with $\mc{B} = \mc{U}\times \mc{S}\times \mc{Z}\times \mc{X}\times \mc{Y} \times \mc{V} $.
\end{theorem}
\vspace{-0.4cm}
\begin{remark}
It is possible to send an additional message $m\in\mc{M}$ with rate $\sf{R}$ corresponding to left-hand side of equation \eqref{eq:Achievability1}, while still satisfying the coordination requirement. This remark also extends to the other results of this article.
\end{remark}

Sketches of proof are available in App. \ref{sec:ProofAchievabilityNC} and  \ref{sec:ProofConverseNC}. 
Achievability result of Theorem \ref{theo:NC} is based on hybrid coding \cite{LimMineroKim(Allerton)10}, \cite{MineroLimKim(ISIT)11} and is stated in \cite{CuffSchieler11}, with a unique auxiliary random variable $W = (W_1,W_2)$, without state informations $S$ and $Z$. Empirical coordination is obtained by considering $(U,S)$ as state information at the encoder and $(Y,Z)$ as a state information at the decoder. This open problem is also related to "state communication" \cite{ChoudhuriKimMitra13} which is solved for Gaussian channels \cite{SutivongChiangCoverKim05}, but remains open for non-causal encoding and decoding. In the following, we connect the duality result of \cite{CoverChiang02} and the separation result of \cite{MerhavShamai03} to empirical coordination.

\section{Optimal Solutions for Particular Cases}\label{sec:OptimalSolutions}

In this section, we characterize the joint probability distributions that are achievable for three particular cases.

\begin{figure}[!h]
\begin{center}
\psset{xunit=0.9cm,yunit=0.9cm}
\begin{pspicture}(0,-0.4)(8.5,1.5)
\pscircle(0,0.5){0.45}
\psframe(2,0)(3,1)
\psframe(7,0)(8,1)
\psline[linewidth=1pt]{->}(0.5,0.5)(2,0.5)
\psline[linewidth=1pt]{->}(3,0.5)(7,0.5)
\psline[linewidth=1pt]{->}(8,0.5)(9,0.5)
\psline[linewidth=1pt]{->}(0,0)(0,-0.5)(2.5,-0.5)(2.5,0)
\psline[linewidth=1pt]{->}(0,1)(0,1.5)(7.5,1.5)(7.5,1)
\rput[u](1,0.8){$U^n$}
\rput[u](1,-0.2){$S^n$}
\rput[u](1,1.8){$Z^n$}
\rput[u](3.75,0.8){$X^n$}
\rput[u](8.5,0.8){$V^n$}
\rput(0,0.5){$\PP_{\sf{usz}}$}
\rput(2.5,0.5){$\C$}
\rput(7.5,0.5){$\D$}
\end{pspicture}
\caption{The perfect channel is defined by $\mc{T}_{\sf{y|xs}} = \UN(y|x)$. }
\label{fig:PerfectChannel}
\end{center}
\end{figure}
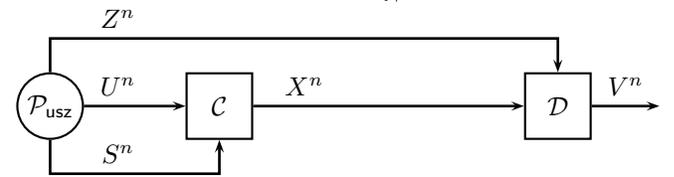

\subsection{Perfect Channel  is defined by $\mc{T}_{\sf{y|xs}} = \UN(y|x)$ as in Fig. \ref{fig:PerfectChannel} }\label{sec:PerfectChannel}


\vspace{-0.6cm}

\begin{theorem}[Perfect Channel]
$\qquad\qquad\qquad\qquad\qquad\qquad\qquad\qquad\qquad\qquad\qquad\qquad\qquad\qquad\qquad\qquad$\label{theo:PerfectChannel}
$1)$ If the joint probability distribution $\QQ(u,s,z,x,v)$ is achievable, then it satisfies:
\begin{small}
\begin{eqnarray}
&\QQ(u,s,z) = \PP_{\sf{usz}}(u,s,z),\quad Z -\!\!\!\!\minuso\!\!\!\!- ( U,S)    -\!\!\!\!\minuso\!\!\!\!-   X.
\end{eqnarray}
\end{small}
$2)$ The probability distribution $\PP_{\sf{usz}}(u,s,z)   \otimes   \QQ(x|u,s) \otimes   \QQ(v|u,s,z,x)$ is achievable if \eqref{eq:PerfectChannel1}. The converse holds. 
\begin{eqnarray}
\max_{{\QQ}\in \Q_{\sf{p}}} \bigg( I(W_2 ; Z|X  ) + H(X)  -  I(X, W_2 ; U, S )   \bigg)  >0 ,  \label{eq:PerfectChannel1}
\end{eqnarray}
$\Q_{\sf{p}}$ is the set of distributions ${\QQ}_{\sf{uszw_2xv}}$ with $W_2$ satisfying marg-\\inals and  $Z -\!\!\!\!\minuso\!\!\!\!- (U , S ) -\!\!\!\!\minuso\!\!\!\!-  (X , W_2 )$ and $V -\!\!\!\!\minuso\!\!\!\!- (X,Z,W_2 ) -\!\!\!\!\minuso\!\!\!\!-  (U,S ) $.
 Support satisfies $|\mc{W}_2| \leq   |\mc{B} |  +1$ and $\QQ \in \Q_{\sf{p}}$ decomposes:
\begin{eqnarray*}
 \PP_{\sf{usz}}(u,s,z)   \otimes   \QQ(x|u,s) \otimes   \QQ( w_2|u,s,x)  \otimes \QQ(v|x,z,w_2).
 \end{eqnarray*}
\end{theorem}
Achievability comes from replacing $Y$ and $W_1$ by $X$ in Theorem \ref{theo:NC} and converse is in App. \ref{sec:ConversePC}. This results extends the coding theorem of Wyner-Ziv \cite{wyner-it-1976} with distortion $d(u,v)$, to the framework of coordination. The message of rate $\textsf{R}$ corresponds  to the channel inputs $X^n$ of rate $\log_2|\mc{X}|$ and the optimal distortion level $D^{\star}$ is obtained by taking the expectation $\E[d(u,v)]$, as in Corollary \ref{coro:RemovingCoordination}. The main difference is that the symbols $(U,S,V)$ are coordinated with $X$. 






\subsection{Lossless Decoding defined by $\QQ(v|u,s,z,x,y) =  \UN(\hat{u}|u)$}\label{sec:losslessDecoding}

The characterization for lossless decoding with correlated source and state is in \cite{LeTreust(CorrelationITW)14}. Achievability is also a particular case of Theorem \ref{theo:NC} by replacing random variables $V$ and $W_2$ by $U$. Joint distribution $\PP_{\sf{usz}}(u,s,z)   \otimes   \QQ(x|u,s) \otimes \mc{T}(y|x,s) \otimes \UN(\hat{u}|u)$ is achievable if  \eqref{eq:LosslessChannel1}. The converse holds. 
\begin{eqnarray}
\max_{{\QQ}\in \Q_{\sf{l}} } \bigg( I(U , W_1 ; Y,Z  )  -  I(W_1 ; S |U )   -  H(U  )  \bigg)  > 0 ,  \label{eq:LosslessChannel1}
\end{eqnarray}
$\Q_{\sf{l}}$ is the set of distributions ${\QQ}_{\sf{uszw_1xy\hat{u}}}$ satisfying marginal conditions and Markov chains $Y -\!\!\!\!\minuso\!\!\!\!- (X , S ) -\!\!\!\!\minuso\!\!\!\!-  (U , Z,W_1 )$ and $Z -\!\!\!\!\minuso\!\!\!\!- (U , S ) -\!\!\!\!\minuso\!\!\!\!-  (X,Y ,W_1 )$. This result extends the coding theorem of Gel'fand-Pinker \cite{gelfand-it-1980} to the framework of coordination. The message is a sequence of source symbols $U$ correlated with the states $(S,Z)$ and channel inputs $X$. Duality between equations \eqref{eq:PerfectChannel1} and \eqref{eq:LosslessChannel1} recalls the duality between channel capacity and rate distortion, as mentioned in \cite{CoverChiang02}.

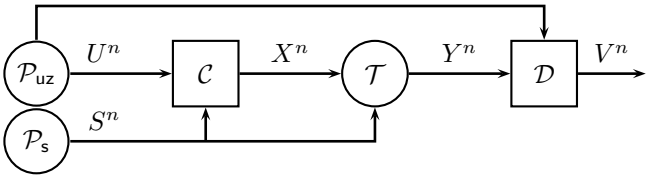
\begin{figure}[!ht]
\begin{center}
\psset{xunit=0.9cm,yunit=0.9cm}
\begin{pspicture}(0,-0.65)(8.5,1.4)
\pscircle(0,0.5){0.44}
\pscircle(0,-0.5){0.44}
\psframe(2,0)(3,1)
\pscircle(5,0.5){0.45}
\psframe(7,0)(8,1)
\psline[linewidth=1pt]{->}(0.5,0.5)(2,0.5)
\psline[linewidth=1pt]{->}(3,0.5)(4.5,0.5)
\psline[linewidth=1pt]{->}(5.5,0.5)(7,0.5)
\psline[linewidth=1pt]{->}(8,0.5)(9,0.5)
\psline[linewidth=1pt]{->}(0.5,-0.5)(5,-0.5)(5,0)
\psline[linewidth=1pt]{->}(2.5,-0.5)(2.5,0)
\psline[linewidth=1pt]{->}(0,1)(0,1.5)(7.5,1.5)(7.5,1)
\rput[u](1,0.8){$U^n$}
\rput[u](1,-0.2){$S^n$}
\rput[u](1,1.75){$Z^n$}
\rput[u](3.75,0.8){$X^n$}
\rput[u](6.25,0.8){$Y^n$}
\rput[u](8.5,0.8){$V^n$}
\rput(0,0.5){$\PP_{\sf{uz}}$}
\rput(0,-0.5){$\PP_{\sf{s}}$}
\rput(2.5,0.5){$\C$}
\rput(5,0.5){$\mc{T}$}
\rput(7.5,0.5){$\D$}
\end{pspicture}
\caption{The random variables of the source $(U,Z,V)$ are independent of the random variables of the channel $(S,X,Y)$.}
\label{fig:MS}
\end{center}
\end{figure}

\subsection{Independent source $(U,Z,V)$, channel $(S,X,Y)$, Fig. \ref{fig:MS}}\label{sec:MS}

\vspace{-0.5cm}

\begin{theorem}[Independent Source and Channel]
\qquad\qquad\qquad
\label{theo:MS}
$1)$ If the product $\QQ(u,z,v)\otimes \QQ(s,x,y)$ of probability distribution  is achievable, then it satisfies:
\begin{footnotesize}
\begin{eqnarray}
\QQ(u,z) = \PP_{\sf{uz}}(u,z),\quad\QQ(s) = \PP_{\sf{s}}(s),\quad \QQ(y|x,s) = \mc{T}(y | x ,s).
\end{eqnarray}
\end{footnotesize}
2) Probability distribution $\PP_{\sf{uz}}(u,z)   \otimes  \QQ(v|u,z) \otimes \PP_{\sf{s}}(s)   \otimes   \QQ(x|s)  \otimes \mc{T}(y | x ,s)$ is achievable if \eqref{eq:Separation1}. The converse holds.
\begin{small}
\begin{eqnarray}
\max_{{\QQ}\in \Q_{\sf{s}}} \bigg( I(W_1 ; Y  )  +  I(W_2 ; Z  )   -  I( W_1 ; S )   -  I( W_2 ; U )    \bigg)  > 0 ,  \label{eq:Separation1}
\end{eqnarray}
\end{small}
$\Q_{\sf{s}}$ is the set of product distributions ${\QQ}_{\sf{uzw_2v}} \otimes {\QQ}_{\sf{sxw_1y}} $
with auxiliary random variables $(W_1,W_2)$ satisfying conditions:
\begin{footnotesize}
\begin{eqnarray*}
Z -\!\!\!\!\minuso\!\!\!\!- U -\!\!\!\!\minuso\!\!\!\!-   W_2 , \quad V -\!\!\!\!\minuso\!\!\!\!- (Z,W_2 ) -\!\!\!\!\minuso\!\!\!\!-  U ,\quad
Y -\!\!\!\!\minuso\!\!\!\!- (X , S ) -\!\!\!\!\minuso\!\!\!\!-  W_1.
\end{eqnarray*}
\end{footnotesize}
Supports of $(W_1,W_2)$ are bounded by $(|\mc{B} |  +1) \cdot(|\mc{B} |  +2)$.
 The product distribution $\QQ \in \Q_{\sf{s}}$ decomposes as follows:
\begin{footnotesize}
\begin{eqnarray*}
 \PP_{\sf{uz}}(u,z) \otimes   \QQ( w_2|u)  \otimes \QQ(v|z,w_2)\otimes   \PP_{\sf{s}}(s)   \otimes \QQ(x,w_1|s) \otimes   \mc{T}(y | x ,s) .
 \end{eqnarray*}
 \end{footnotesize}
\end{theorem}

As mentioned in \cite{LeTreust(EmpiricalCoordination)14}, the independence between the probability  distributions of the source and channel induces the separation of the source coding and the channel coding. This result is related to Theorem 1 in \cite{MerhavShamai03}, with expected distortion  $d(u,v)$ and channel cost $c(x)$ functions. By considering $(U,Z,W_2,V)$ independent of $(S,X,W_1,Y)$ and introducing three indexes $(m,l,j)$ with rates $(\textsf{R}_{m} ,\textsf{R}_{l} ,\textsf{R}_{j} )$ satisfying,
\begin{small}
\begin{eqnarray*}
\textsf{R}_{m} + \textsf{R}_l \geq& I(W_2;U),\quad \qquad\qquad  \textsf{R}_l & \leq  I(W_2;Z),\\
\textsf{R}_{j}  \geq &I(W_1;S) , \; \quad \quad \textsf{R}_{m} + \textsf{R}_j & \leq  I(W_1;Y),
\end{eqnarray*}
\end{small}
the achievability of Theorem \ref{theo:NC} becomes a separated coding scheme.
The author would like to thank Pablo Piantanida, Matthieu Bloch and Claudio Weidmann for useful discussions about the independence of the auxiliary random variables $(W_1,W_2)$ in the converse proof of Theorem \ref{theo:MS}.


\section{Causal Encoding and Decoding}\label{sec:CausalEncDec}

The encoding (resp. decoding) is causal if for all $i \in \{1,\ldots,n\}$, we have $X_i = f_i(U^{i},S^{i})$, (resp. $V_i = g_i(Y^{i},Z^{i})$). We characterize of the set of achievable joint probability distributions for causal encoding and for causal decoding.

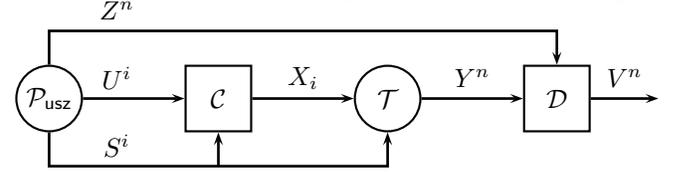
\begin{figure}[!ht]
\begin{center}
\psset{xunit=0.9cm,yunit=0.9cm}
\begin{pspicture}(0,-0.4)(8.5,1.5)
\pscircle(0,0.5){0.45}
\psframe(2,0)(3,1)
\pscircle(5,0.5){0.45}
\psframe(7,0)(8,1)
\psline[linewidth=1pt]{->}(0.5,0.5)(2,0.5)
\psline[linewidth=1pt]{->}(3,0.5)(4.5,0.5)
\psline[linewidth=1pt]{->}(5.5,0.5)(7,0.5)
\psline[linewidth=1pt]{->}(8,0.5)(9,0.5)
\psline[linewidth=1pt]{->}(0,0)(0,-0.5)(5,-0.5)(5,0)
\psline[linewidth=1pt]{->}(2.5,-0.5)(2.5,0)
\psline[linewidth=1pt]{->}(0,1)(0,1.5)(7.5,1.5)(7.5,1)
\rput[u](1,0.8){$U^i$}
\rput[u](1,-0.2){$S^i$}
\rput[u](1,1.8){$Z^{n}$}
\rput[u](3.75,0.8){$X_i$}
\rput[u](6.25,0.8){$Y^{n}$}
\rput[u](8.5,0.8){$V^n$}
\rput(0,0.5){$\PP_{\sf{usz}}$}
\rput(2.5,0.5){$\C$}
\rput(5,0.5){$\mc{T}$}
\rput(7.5,0.5){$\D$}
\end{pspicture}
\caption{Causal encoding function $f_i: \mc{U}^i \times \mc{S}^i \rightarrow \mc{X}$, for all $i\in\{1,\ldots,n\}$.}
\end{center}
\end{figure}

\vspace{-0.3cm}

\begin{theorem}[Causal Encoding and Non-Causal Decoding] 
$\qquad\qquad\qquad\qquad\qquad\qquad\qquad\qquad\qquad\qquad\qquad\qquad\qquad\qquad\qquad$\label{theo:CE}
$1)$ If $\QQ(u,s,z,x,y,v)$ is achievable, then it satisfies:
\begin{small}
\begin{eqnarray}
\begin{cases}
&\QQ(u,s,z) = \PP_{\sf{usz}}(u,s,z),\quad  \QQ(y|x,s) = \mc{T}(y | x ,s),\\
&Y -\!\!\!\!\minuso\!\!\!\!- ( X,S)    -\!\!\!\!\minuso\!\!\!\!-   (U,Z) , \quad Z -\!\!\!\!\minuso\!\!\!\!- ( U,S)    -\!\!\!\!\minuso\!\!\!\!-   (X,Y).
\end{cases}
\end{eqnarray}
\end{small}
2) The distribution $\PP_{\sf{usz}}(u,s,z)   \otimes   \QQ(x|u,s)  \otimes \mc{T}(y | x ,s)  \otimes \QQ(v|u,s,z,x,y) $ is achievable if \eqref{eq:AchievabilityCE1}. The converse holds.
\begin{eqnarray}
\max_{{\QQ}\in \Q_{\sf{e}}} \bigg( I(W_1,W_2 ; Y, Z  )   -  I( W_2 ; U, S | W_1 )   \bigg)  > 0 ,  \label{eq:AchievabilityCE1}
\end{eqnarray}
$\Q_{\sf{e}}$ is the set of joint distributions ${\QQ}_{\sf{uszw_1w_2xyv}}$ with auxiliary random variables $(W_1 , W_2)$ satisfying marginal cond. and:
\begin{small}
\begin{eqnarray*}
\begin{cases}
(U,S) \text{ independent of } W_1,\qquad
X -\!\!\!\!\minuso\!\!\!\!- (U , S ,W_1) -\!\!\!\!\minuso\!\!\!\!-  W_2 ,\\ 
Y -\!\!\!\!\minuso\!\!\!\!- (X , S ) -\!\!\!\!\minuso\!\!\!\!-  (U,Z, W_1,W_2 ) ,\\
Z -\!\!\!\!\minuso\!\!\!\!- (U , S ) -\!\!\!\!\minuso\!\!\!\!-  (X , Y, W_1, W_2 ) ,\\
V -\!\!\!\!\minuso\!\!\!\!- (Y,Z,W_1, W_2 ) -\!\!\!\!\minuso\!\!\!\!-  (U,S ,X) .
\end{cases}
\end{eqnarray*}
\end{small}
Supports of $(W_1,W_2)$ are bounded by $ (|\mc{B} |  +1) \cdot(|\mc{B} |  +2)$. 
The probability distribution $\QQ \in \Q_{\sf{e}}$ decomposes as follows:
\begin{eqnarray*}
\PP_{\sf{usz}}   \otimes   \QQ_{\sf{w_1}} \otimes   \QQ_{\sf{w_2|usw_1}} 
\otimes \QQ_{\sf{x|usw_1}}  \otimes   \mc{T}_{\sf{y|xs}}   \otimes  \QQ_{\sf{v|yzw_1w_2}}.
 \end{eqnarray*}
\end{theorem}

Proofs are available in App. \ref{sec:ProofAchievabilityNC} and  \ref{sec:ProofConverseNC}. This result is a particular case of Theorem 2 in \cite{CuffSchieler11}, by considering  $(U,S)$ as information source and $(Y,Z)$ as channel output. Empirical coordination is equivalent to  joint source channel coding with two-sided state information and correlated source and state. Theorem \ref{theo:CE} also reduces to Theorem 3 in \cite{ChoudhuriKimMitra13}, by considering $(U,S)$ as channel state and $(Y,Z)$ as channel output.  


When considering strictly causal encoding instead of causal encoding, the optimal solution is obtained by replacing $W_1$ by $X$ in equation \eqref{eq:AchievabilityCE1}.

\begin{corollary}[Causal Encoding without Coordination]$\qquad\qquad\qquad\qquad$\label{coro:RemovingCoordination}
Consider a distortion function 
$d:\mc{U} \times \mc{V} \mapsto \R$. The distortion level $D\geq0$ is achievable if and only if:
\small
\begin{eqnarray*}
\max_{ \QQ_{\sf{w_1}} ,   \QQ_{\sf{w_2|usw_1}}  ,  \QQ_{\sf{x|usw_1}}, \atop  \QQ_{\sf{v|yzw_1w_2}},  \E [d(U,V)] \leq D } \bigg( I(W_1,W_2 ; Y, Z  )   -  I( W_2 ; U, S | W_1 )   \bigg)  \geq 0 ,  \label{eq:coro1}
\end{eqnarray*}
\end{corollary}

Corollary \ref{coro:RemovingCoordination} is a direct consequence of Theorem \ref{theo:CE}. The distortion level $D \geq0$ is achievable if and only if there exists a joint distribution $\PP_{\sf{usz}}   \otimes   \QQ_{\sf{w_1}} \otimes   \QQ_{\sf{w_2|usw_1}}  \otimes \QQ_{\sf{x|usw_1}}  \otimes   \mc{T}_{\sf{y|xs}}   \otimes  \QQ_{\sf{v|yzw_1w_2}}$ that satisfies \eqref{eq:coro1}, such that $\E [d(U,V)]  \leq D$. Similar analysis is mentioned in  \cite{LeTreust(EmpiricalCoordination)14}  for the proof of Theorem V.3.

Causal decoding and non-causal encoding is considered in \cite{LeTreust(EmpiricalCoordination)14} without $S$ and $Z$. Joint probability distribution $\PP_{\sf{usz}}  \otimes   \QQ_{\sf{x|us}}  \otimes \mc{T}_{\sf{y|xs}}  \otimes \QQ_{\sf{v|uszxy}}$
 is achievable if \eqref{eq:CD1}. Converse holds.
\begin{eqnarray}
\max_{\QQ_{\sf{xw_1w_2|us}} ,\atop \QQ_{\sf{v|yzw_2}}} \bigg( I(W_1 ; Y, Z | W_2  )   -  I( W_1,W_2 ; U, S )   \bigg)  > 0 ,  \label{eq:CD1}
\end{eqnarray}
Remark that $V$ depends on $(Y,Z,W_2)$ but not on $W_1$. When considering strictly causal decoding instead of causal decoding, the optimal solution is obtained by replacing $W_2$ by $V$ in equation \eqref{eq:CD1}. More details are provided in \cite{LeTreust(EmpiricalCoordination)14}.


\section{Conclusion}\label{sec:Conclusion}

The problem of empirical coordination is closely related to the joint source-channel coding with two-sided state information with correlation between source and state. These two problems are also related to state communication. We provide achievability and converse results for the non-causal case, that is open. We characterize the optimal solutions for perfect channel, for lossless decoding, for independent source and channel, for causal encoding and for causal decoding.


\appendix\label{sec:App}

The full versions of the proofs are stated in \cite{LeTreust(InternalRapportISIT)15}.

\subsection{Sketch of Achievability of Theorem \ref{theo:NC}}\label{sec:ProofAchievabilityNC}

Consider $\QQ(u,s,z,w_1,w_2,x,y,v)\in\Q$ that achieves the maximum in \eqref{eq:Achievability1}. There exists $\delta>0$ and rate $\sf{R}\geq 0$ such that:
\begin{eqnarray} 
\sf{R} &\geq&  I(W_1,W_2 ; U , S ) + \delta ,\label{eq:AchievabilityNC1}\\
\sf{R} &\leq&  I( W_1, W_2 ; Y , Z ) - \delta  .\label{eq:AchievabilityNC2}
\end{eqnarray}

\begin{itemize}
\item[$\bullet$] \textit{Random codebook.} We generate $|\mc{M} |= 2^{n    \sf{R}} $ pairs of sequences $(W_1^n(m) , W_2^n(m))$ with index $m\in\mc{M}$ drawn from the i.i.d. marginal probability distribution $\QQ_{\sf{w_1w_2}}^{\otimes n}$. 
\item[$\bullet$] \textit{Encoding function.}  Encoder finds the index $m\in \mc{M}$ such that $(U^n,S^n, W_1^n(m) , W_2^n(m) ) \in A_{\varepsilon}^{{\star}{n}}(\QQ)$ are jointly typical. It sends $X^n$ drawn from $\QQ_{\sf{x|usw_1w_2}}^{\otimes n}$ depending on $(U^n, S^n, W_1^n(m) , W_2^n(m) )$.
\item[$\bullet$] \textit{Decoding function.} Decoder finds the index $m \in \mc{M} $  such that $(Y^n,Z^n, W_1^n(m) , W_2^n(m) ) \in A_{\varepsilon}^{{\star}{n}}(\QQ)$ are jointly typical. Decoder returns $V^n$ drawn from $\QQ_{\sf{v|yzw_1w_2}}^{\otimes n}$ depending on $(Y^n, Z^n, W_1^n(m) , W_2^n(m) )$.
\end{itemize}
The pair $(W_1,W_2)$ can be replaced by a single $W$. From properties of typical sequences, packing and covering Lemmas stated in \cite{ElGammalKim(book)11} pp. 27, 46 and 208, equations \eqref{eq:AchievabilityNC1}, \eqref{eq:AchievabilityNC2} imply that the expected probability of error is bounded for all $n\geq\bar{n}$:
\begin{tiny}
\begin{eqnarray}
&&\E_c\bigg[ \PP\bigg((U^n,S^n) \notin A_{\varepsilon}^{{\star}{n}}(\QQ) \bigg)\bigg]  \leq \varepsilon,\\
&&\E_c\bigg[ \PP\bigg( \forall  m \in \mc{M}  ,\quad (U^n,S^n, W_1^n(m) , W_2^n(m) ) \notin A_{\varepsilon}^{{\star}{n}}(\QQ) \bigg)\bigg]  \leq \varepsilon,\\
&&\E_c\bigg[ \PP\bigg(  \exists m'\neq  m  ,\text{ s.t. } (Y^n,Z^n, W_1^n(m') , W_2^n(m') ) \in A_{\varepsilon}^{{\star}{n}} (\QQ)    \bigg)\bigg]   \leq \varepsilon.
\end{eqnarray}
\end{tiny}
There exists a code $c^{\star} \in \mc{C}(n)$ such that sequences are jointly typical for distribution $\PP_{\sf{usz}} \otimes   \QQ_{\sf{xw_1w_2|us}} \otimes  \mc{T}_{\sf{y|xs}}   \otimes  \QQ_{\sf{v|yzw_1w_2}}$ with probability more than $1 - 3\varepsilon$.

%


\subsection{Sketch of Converse of Theorem \ref{theo:NC}}\label{sec:ProofConverseNC}

Consider code $c(n) \in \mc{C}$ with small error probability $\PP_{\sf{e}}(c)$.  
\begin{scriptsize}
\begin{eqnarray}
0&\leq& \sum_{i=1}^n I( U_{i+1}^n , S_{i+1}^n , Y_{i+1}^n , Z_{i+1}^n  ; Y_i, Z_i | Y^{i-1} , Z^{i-1})   \nonumber\\
&-&  \sum_{i=1}^n I( Y^{i-1} , Z^{i-1}  ; U_i, S_i  |  U_{i+1}^n , S_{i+1}^n , Y_{i+1}^n , Z_{i+1}^n  ) \label{eq:ConverseNC2}\\
&=& \sum_{i=1}^n I( W_{1,i}  ; Y_i, Z_i |  W_{2,i} )    -  \sum_{i=1}^n I(  W_{2,i}   ; U_i, S_i  |  W_{1,i}  )
\label{eq:ConverseNC3} \\
&\leq& n \cdot \max_{ \QQ \in\Q}\bigg(   I( W_{1}  ; Y, Z |  W_{2} )    -   I(  W_{2}   ; U, S |  W_{1} ) \bigg) .\label{eq:ConverseNC7}
\end{eqnarray}
\end{scriptsize}
Eq. \eqref{eq:ConverseNC2} is due to Csisz\'{a}r Sum Identity and properties of MI.\\
Eq. \eqref{eq:ConverseNC3} introduces auxiliary random variables $ W_{1,i}  = ( U_{i+1}^n , S_{i+1}^n , Y_{i+1}^n , Z_{i+1}^n )$ and $ W_{2,i}  = ( Y^{i-1} , Z^{i-1} )$ satisfying:
\begin{eqnarray}
&&Y_i -\!\!\!\!\minuso\!\!\!\!- (X_i , S_i ) -\!\!\!\!\minuso\!\!\!\!-  (U_i,Z_i, W_{1,i},W_{2,i} ) , \label{eq:MarkovNC1}\\
&&Z_i -\!\!\!\!\minuso\!\!\!\!- (U_i , S_i ) -\!\!\!\!\minuso\!\!\!\!-  (X_i , Y_i, W_{1,i},W_{2,i} ) ,\label{eq:MarkovNC2} \\
&&V_i -\!\!\!\!\minuso\!\!\!\!- (Y_i,Z_i,W_{1,i},W_{2,i} ) -\!\!\!\!\minuso\!\!\!\!-  (U_i,S_i,X_i ) .\label{eq:MarkovNC3} 
\end{eqnarray}
This is due to memoryless channel, i.i.d. source and non-causal decoding since $(Y^n,Z^n)$ is included in $(Y_i,Z_i,W_{1,i},W_{2,i} )$.\\
Eq. \eqref{eq:ConverseNC7} comes from taking the maximum over the set $\Q$.


\subsection{Sketch of Converse of Theorem \ref{theo:PerfectChannel}}\label{sec:ConversePC}
 
Consider code $c(n) \in \mc{C}$ with small error probability $\PP_{\sf{e}}(c)$.  
\begin{scriptsize}
\begin{eqnarray}
0&=& H(X^n,Z^n ) -  I(X^n,Z^n ; U^n,S^n)  - H( X^n,Z^n | U^n,S^n )\label{eq:ConversePC1} \\
&\leq&  \sum_{i=1}^n  H(X_i,Z_i  ) - \sum_{i=1}^n  I(X^n,Z^n ; U_i,S_i | U_{i+1}^n , S_{i+1}^n  )  \nonumber\\
&- &  H( X^n,Z^n | U^n,S^n  )\label{eq:ConversePC2} \\
&=& \sum_{i=1}^n I(X^n,Z^n , U_{i+1}^n , S_{i+1}^n ; X_i,Z_i  )  \nonumber\\
&- & \sum_{i=1}^n I(X^n,Z^n , U_{i+1}^n , S_{i+1}^n  ; U_i,S_i  )  - H( X^n,Z^n | U^n,S^n  ) \label{eq:ConversePC3} \\
&\leq& \sum_{i=1}^n I(X^n,Z^{-i} , U_{i+1}^n , S_{i+1}^n ; X_i,Z_i  ) \nonumber\\
&- & \sum_{i=1}^nI(X^n,Z^{-i} , U_{i+1}^n , S_{i+1}^n  ; U_i,S_i  ) \label{eq:ConversePC7}  \\
&=& \sum_{i=1}^n I(X_i,W_{2,i} ; X_i,Z_i  ) -  I(X_i,W_{2,i} ; U_i,S_i  )   \label{eq:ConversePC8}  \\
&\leq& n \cdot \max_{ \QQ \in\Q_{\sf{p}} } \bigg( I(X,W_{2} ; X,Z ) -  I(X,W_{2} ; U,S)  \bigg) \label{eq:ConversePC12} .
\end{eqnarray}
\end{scriptsize}
Eq. \eqref{eq:ConversePC1}, \eqref{eq:ConversePC2} are due to properties of mutual information.\\
Eq. \eqref{eq:ConversePC3} is due to the i.i.d. source $(U,S)$.\\
Eq. \eqref{eq:ConversePC7}  is due to the i.i.d. source $(U,S,Z)$: 
$ H(Z^n | U^n,S^n)  = \sum H(Z_i   | U_i,S_i  ) = \sum H(Z_i   | X^n,Z^{-i} , U_{i+1}^n , S_{i+1}^n ,U_i,S_i  ) $.\\
Eq. \eqref{eq:ConversePC8} introduces auxiliary random variable $ W_{2,i}  = ( X^{-i}, Z^{ - i},  U_{i+1}^n , S_{i+1}^n  )$ that satisfies Markov chains: $Z_i -\!\!\!\!\minuso\!\!\!\!- (U_i , S_i ) -\!\!\!\!\minuso\!\!\!\!-  (X_i , W_{2,i} )$ and $V_i -\!\!\!\!\minuso\!\!\!\!- (X_i,Z_i,W_{2,i} ) -\!\!\!\!\minuso\!\!\!\!-  (U_i,S_i )$.
This is due to i.i.d. source $(U,S,Z)$ and non-causal decoding.\\
Eq. \eqref{eq:ConversePC12} comes from taking the maximum over the set $\Q_{\sf{p}}$.


\subsection{Sketch of Achievability of Theorem \ref{theo:CE}}\label{app:theoCE}

Consider $\QQ \in \Q_{\sf{e}}$ that maximizes \eqref{eq:AchievabilityCE1} and 
block-Markov code $c\in \mc{C}(n)$ over $B\in \N$ blocs of length $n\in \N$ with:
\begin{eqnarray}
\textsf{R}  & \geq&  I(  W_2   ; U,S  | W_1    )  + \delta   , \label{eq:AchCausalEncoding1}\\
\textsf{R}   & \leq&  I( W_1  ; Y ,Z) + I( W_2  ; Y ,Z | W_1)  -  \delta . \label{eq:AchCausalEncoding2}
\end{eqnarray}
\begin{itemize}
\item[$\bullet$] \textit{Random codebook.} We generate $| \mc{M}  |= 2^{n   \sf{R}  } $ sequences $W_1^n(m)$  drawn from $\QQ_{\sf{w}_1}^{\otimes n} $ with index  $m\in \mc{M} $. For each $m\in \mc{M} $, we generate the same number $| \mc{M}  |= 2^{n   \sf{R} } $ of sequences $ W_2^n(m,\hat{m})$ with index  $\hat{m} \in \mc{M} $, drawn from $\QQ_{\sf{w_2|w_1}}^{\otimes n} $ depending on $W_1^n(m)$.  
\item[$\bullet$] \textit{Encoding function.} It recalls $m_{b-1}$ and finds $ m_b\in \mc{M}$ s.t. sequences  $(U^n_{b-1},S^n_{b-1}, W^n_{1}(m_{b-1}) ,W_{2}^n(m_{b-1},m_b) )\in A_{\varepsilon}^{{\star}{n}}(\QQ)$ are jointly typical in block $b-1$. It deduces $W_{1}^n(m_b) $ for block $b$ and sends $X_b^n$ drawn from $\QQ_{\sf{x|usw_1}}^{\otimes n}$  depending on $(U^n_b ,S^n_b ,W_{1}^n(m_b) )$.
\item[$\bullet$] \textit{Decoding function.} It recalls $m_{b-1}$ and finds $m_b \in \mc{M}$ such that  $(Y^n_{b-1} , Z^n_{b-1}  ,W^n_{1}(m_{b-1}), W^n_{2}(m_{b-1} , m_b) ) \in A_{\varepsilon}^{{\star}{n}}( \QQ)$ and $(Y^n_b , Z^n_b  ,W^n_{1}(m_b) ) \in A_{\varepsilon}^{{\star}{n}}( \QQ)$ are jointly typical. It returns $V^n_{b-1}$ drawn from $\QQ_{\sf{v|yzw_1w_2}}^{\otimes n}$  depending on $(Y^n_{b-1} , Z^n_{b-1} , W^n_{1}(m_{b-1}), W^n_{2}(m_{b-1} ,  m_b))$.
\end{itemize}
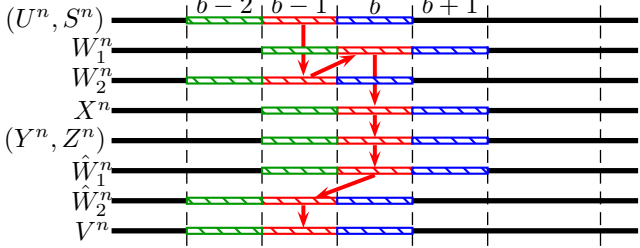
\begin{figure}[!ht]
\begin{center}
\psset{xunit=0.5cm,yunit=0.2cm}
\begin{pspicture}(-2,-5.2)(12.5,7.35)
\psline[linewidth=2pt](0,-6)(14,-6)
\psline[linewidth=2pt](0,-4)(14,-4)
\psline[linewidth=2pt](0,-2)(14,-2)
\psline[linewidth=2pt](0,0)(14,0)
\psline[linewidth=2pt](0,2)(14,2)
\psline[linewidth=2pt](0,4)(14,4)
\psline[linewidth=2pt](0,6)(14,6)
\psline[linewidth=2pt](0,8)(14,8)
\psline[linewidth=0.5pt,linestyle=dashed ](2,-7)(2,9)
\psline[linewidth=0.5pt,linestyle=dashed ](4,-7)(4,9)
\psline[linewidth=0.5pt,linestyle=dashed ](6,-7)(6,9)
\psline[linewidth=0.5pt,linestyle=dashed ](8,-7)(8,9)
\psline[linewidth=0.5pt,linestyle=dashed ](10,-7)(10,9)
\psline[linewidth=0.5pt,linestyle=dashed ](13,-7)(13,9)
\rput[u](-1.5,8){$(U^n,S^n)$}
\rput[u](-0.5,6){$W_1^n$}
\rput[u](-0.5,4){$W_2^n$}
\rput[u](-0.5,2){$X^n$}
\rput[u](-1.5,0){$(Y^n,Z^n)$}
\rput[u](-0.5,-2){$\hat{W}_1^n$}
\rput[u](-0.5,-4){$\hat{W}_2^n$}
\rput[u](-0.5,-6){${V}^n$}
\rput[u](3,9){$b-2$}
\rput[u](5,9){$b-1$}
\rput[u](7,9){$b$}
\rput[u](9,9){$b+1$}
\pspolygon[fillstyle=vlines*,linecolor=red,hatchcolor = red](4,7.8)(4,8.2)(6,8.2)(6,7.8)
\pspolygon[fillstyle=vlines*,linecolor=red,hatchcolor = red](6,5.8)(6,6.2)(8,6.2)(8,5.8)
\pspolygon[fillstyle=vlines*,linecolor=red,hatchcolor = red](4,3.8)(4,4.2)(6,4.2)(6,3.8)
\pspolygon[fillstyle=vlines*,linecolor=red,hatchcolor = red](6,1.8)(6,2.2)(8,2.2)(8,1.8)
\pspolygon[fillstyle=vlines*,linecolor=red,hatchcolor = red](6,-0.2)(6,0.2)(8,0.2)(8,-0.2)
\pspolygon[fillstyle=vlines*,linecolor=red,hatchcolor = red](6,-2.2)(6,-1.8)(8,-1.8)(8,-2.2)
\pspolygon[fillstyle=vlines*,linecolor=red,hatchcolor = red](4,-4.2)(4,-3.8)(6,-3.8)(6,-4.2)
\pspolygon[fillstyle=vlines*,linecolor=red,hatchcolor = red](4,-6.2)(4,-5.8)(6,-5.8)(6,-6.2)
\psline[linewidth=1.5pt,linecolor=red  ]{->}(7,5.7)(7,2.3)
\psline[linewidth=1.5pt,linecolor=red ]{->}(7,1.7)(7,0.2)
\psline[linewidth=1.5pt,linecolor=red ]{->}(7,-0.3)(7,-1.8)
\psline[linewidth=1.5pt,linecolor=red ]{->}(7,-2.3)(5.4,-3.8)
\psline[linewidth=1.5pt,linecolor=red ]{->}(5.1,-4.3)(5.1,-5.8)
\pspolygon[fillstyle=vlines*,linecolor=blue,hatchcolor = blue](6,7.8)(6,8.2)(8,8.2)(8,7.8)
\pspolygon[fillstyle=vlines*,linecolor=blue,hatchcolor = blue](8,5.8)(8,6.2)(10,6.2)(10,5.8)
\pspolygon[fillstyle=vlines*,linecolor=blue,hatchcolor = blue](6,3.8)(6,4.2)(8,4.2)(8,3.8)
\pspolygon[fillstyle=vlines*,linecolor=blue,hatchcolor = blue](8,1.8)(8,2.2)(10,2.2)(10,1.8)
\pspolygon[fillstyle=vlines*,linecolor=blue,hatchcolor = blue](8,-0.2)(8,0.2)(10,0.2)(10,-0.2)
\pspolygon[fillstyle=vlines*,linecolor=blue,hatchcolor = blue](8,-2.2)(8,-1.8)(10,-1.8)(10,-2.2)
\pspolygon[fillstyle=vlines*,linecolor=blue,hatchcolor = blue](6,-4.2)(6,-3.8)(8,-3.8)(8,-4.2)
\pspolygon[fillstyle=vlines*,linecolor=blue,hatchcolor = blue](6,-6.2)(6,-5.8)(8,-5.8)(8,-6.2)
\psline[linewidth=1.5pt,linecolor=red  ]{->}(5.1,7.7)(5.1,4.3)
\psline[linewidth=1.5pt,linecolor=red  ]{<-}(6.5,5.7)(5.3,4.3)
\pspolygon[fillstyle=vlines*,linecolor=vert,hatchcolor = vert](2,7.8)(2,8.2)(4,8.2)(4,7.8)
\pspolygon[fillstyle=vlines*,linecolor=vert,hatchcolor = vert](4,5.8)(4,6.2)(6,6.2)(6,5.8)
\pspolygon[fillstyle=vlines*,linecolor=vert,hatchcolor = vert](2,3.8)(2,4.2)(4,4.2)(4,3.8)
\pspolygon[fillstyle=vlines*,linecolor=vert,hatchcolor = vert](4,1.8)(4,2.2)(6,2.2)(6,1.8)
\pspolygon[fillstyle=vlines*,linecolor=vert,hatchcolor = vert](4,-0.2)(4,0.2)(6,0.2)(6,-0.2)
\pspolygon[fillstyle=vlines*,linecolor=vert,hatchcolor = vert](4,-2.2)(4,-1.8)(6,-1.8)(6,-2.2)
\pspolygon[fillstyle=vlines*,linecolor=vert,hatchcolor = vert](2,-4.2)(2,-3.8)(4,-3.8)(4,-4.2)
\pspolygon[fillstyle=vlines*,linecolor=vert,hatchcolor = vert](2,-6.2)(2,-5.8)(4,-5.8)(4,-6.2)
\end{pspicture}
\end{center}
\label{fig:AchCausalEncoding}
\end{figure}
Equations \eqref{eq:AchCausalEncoding1}, \eqref{eq:AchCausalEncoding2} imply for all $n\geq\bar{n}$ and for a large number of blocks $B\in \N$, the  sequences are jointly typical with large probability:
\begin{tiny}
\begin{eqnarray*}
&&\E_c\bigg[ \PP\bigg((U^n,S^n) \notin A_{\varepsilon}^{{\star}{n}}(\QQ) \bigg)\bigg]  \leq \varepsilon,\\
&&\E_c\bigg[ \PP\bigg( \forall  m \in \mc{M}  , 
(U^n_{b-1},S^n_{b-1}, W^n_{1}(m_{b-1}) ,W_{2}^n(m_{b-1},m) ) \notin A_{\varepsilon}^{{\star}{n}}(\QQ) \bigg)\bigg]  \leq \varepsilon, \\
&&\E_c\bigg[ \PP\bigg(  \exists m'\neq  m  ,\text{ s.t. } 
\Big\{ (Y^n_b , Z^n_b  ,W^n_{1}(m') ) \in A_{\varepsilon}^{{\star}{n}}( \QQ)\Big\} \cap \nonumber \\
&& \qquad \qquad \qquad \Big\{(Y^n_{b-1} , Z^n_{b-1}  ,W^n_{1}(m_{b-1}), W^n_{2}(m_{b-1} , m') ) \in A_{\varepsilon}^{{\star}{n}}( \QQ) \Big\} \bigg)\bigg]   \leq \varepsilon.\label{eq:achievDouble}
\end{eqnarray*}
\end{tiny}


\subsection{Sketch of Converse of Theorem \ref{theo:CE}}\label{app:theoCEconv}
Consider code $c(n) \in \mc{C}$ with small error probability  $\PP_{\sf{e}}(c)$.
\begin{scriptsize}
\begin{eqnarray}
0&\leq&  \sum_{i=1}^n I(   U^{i-1}   , S^{i-1} , Y^n_{i+1} , Z^n_{i+1} ;  Y_i , Z_i  )  \nonumber \\
&-&   \sum_{i=1}^n I(Y^n_{i+1}  Z^n_{i+1} ;  U_i  ,S_i |  U^{i-1} , S^{i-1} )  \label{eq:ConvCausalEncoding2} \\
&=&  \sum_{i=1}^n I(  W_{1,i}   ,W_{2,i}    ;  Y_i  , Z_i  )   -   \sum_{i=1}^n I( W_{2,i}   ;  U_i ,S_i  |  W_{1,i} )     \label{eq:ConvCausalEncoding4} \\
&\leq &n \cdot  \max_{{\QQ}\in \Q_{\sf{e}}}   \bigg(   I(   W_1  , W_{2}     ;  Y ,Z)     - I( W_{2}  ;  U ,S | W_1 )  \bigg)  \label{eq:ConvCausalEncoding7}.
\end{eqnarray}
\end{scriptsize}
Eq. \eqref{eq:ConvCausalEncoding2} is due to Csisz\'{a}r Sum Identity and properties of MI.\\
Eq. \eqref{eq:ConvCausalEncoding4} introduces auxiliary random variables $ W_{1,i} =  (U^{i-1} , S^{i-1})$ and $ W_{2,i} =  ( Y^n_{i+1} , Z^n_{i+1} )$ satisfying:
\begin{scriptsize}
\begin{eqnarray}
&& (U_i,S_i) \text{ are independent of } W_{1,i}, \label{eq:ConvCEMarkov1} \\
&& X_i -\!\!\!\!\minuso\!\!\!\!-   (U_i , S_i, W_{1,i} ) -\!\!\!\!\minuso\!\!\!\!-  W_{2,i} , \label{eq:ConvCEMarkov2}  \\
&& Y_i -\!\!\!\!\minuso\!\!\!\!-   (X_i,S_i) -\!\!\!\!\minuso\!\!\!\!-  (U_i , Z_i,W_{1,i},W_{2,i}) , \label{eq:ConvCEMarkov3}  \\
&& Z_i -\!\!\!\!\minuso\!\!\!\!-   (U_i,S_i) -\!\!\!\!\minuso\!\!\!\!-  (X_i , Y_i,W_{1,i},W_{2,i}) , \label{eq:ConvCEMarkov4}  \\
&& V_i -\!\!\!\!\minuso\!\!\!\!-   (Y_i, Z_i, W_{1,i}, W_{2,i}) -\!\!\!\!\minuso\!\!\!\!-  (U_{i} ,S_i,X_i). \label{eq:ConvCEMarkov5} 
\end{eqnarray}
\end{scriptsize}
Eq. \eqref{eq:ConvCausalEncoding7} comes from taking the maximum over the set $\Q_{\sf{e}}$.





\end{document}